\documentclass[11pt,14paper]{article}
\usepackage{amssymb}
\usepackage{graphicx}
\usepackage{amsmath}
\usepackage{url}
\usepackage{fullpage}
\usepackage{authblk}
\usepackage{natbib}

\def\sealevel{sea-level~}

\def\rate{$\textrm{mm yr}^{-1}$~}
\def\rates{$\textrm{mm yr}^{-1}$}

\title{New estimates of ongoing sea level change and land movements caused by Glacial Isostatic Adjustment in the Mediterranean region}
\author[1]{Giorgio Spada}
\author[2]{Daniele Melini}
\affil[1]{Dipartimento di Fisica e Astronomia (DIFA), Alma Mater Studiorum Universit\`a di Bologna, Italia}
\affil[2]{Istituto Nazionale di Geofisica e Vulcanologia (INGV), Sezione di Sismologia e Tettonofisica, Roma, Italia}
\date{}

\begin{document}

\linespread{1.4}

\maketitle

\textit{This is a pre-copyedited, author-produced PDF of an article accepted for publication in Geophysical Journal International following peer review. The version of record [G. Spada, D. Melini, New estimates of ongoing sea level change and land movements caused by Glacial Isostatic Adjustment in the Mediterranean region, Geophysical Journal International, Volume 229, Issue 2, May 2022, Pages 984–998] is available online at: }\url{https://doi.org/10.1093/gji/ggab508}

\begin{abstract}
Glacial Isostatic Adjustment (GIA) caused by the melting of past {ice sheets is still a major cause of}  \sealevel variations and 3-D crustal deformation in the 
Mediterranean region. However, since the contribution of GIA cannot be separated from those of 
oceanic or tectonic origin, its role can be only assessed by numerical modelling, {solving} the gravitationally self-consistent Sea Level Equation. Nonetheless, uncertainties about 
the melting history of the late-Pleistocene ice sheets and the rheological profile of the Earth's 
mantle affect the GIA predictions by an unknown amount. Estimating the GIA modelling uncertainties 
would be particularly important in the Mediterranean region, due to the amount of high quality 
geodetic data from space-borne {and} ground-based observations currently available, whose 
interpretation demands a suitable isostatic correction. Here we first review previous results
about the effects of GIA in the Mediterranean Sea, enlightening the variability of all the fields 
affected by the persistent condition of isostatic disequilibrium. Then, for the first time in this 
region, we adopt an ensemble modelling approach {to better constrain} the present-day GIA {contributions} to \sealevel rise and geodetic variations, and {their} uncertainty.

\end{abstract}

\section{Introduction}

Following the seminal works of \citet{flemming1978holocene} and \citet{pirazzoli2005review}, the history of relative 
\sealevel across the Mediterranean Sea during the last millennia has been the subject of a number of investigations. Often, these have 
been focused on specific areas from which paleo \sealevel indicators are available, based upon geological, 
geomorphological and archaeological evidence during the last millennia~\citep[see][and references therein]{lambeck1995late,lambeck2004sea-ii,lambeck2004sea-i,sivan2001holocene,antonioli2009holocene,evelpidou2012late,mauz2015terminal,vacchi2016multiproxy,vacchi2018new}. 
However, the reconstruction of the history of sea level since the Last Glacial Maximum is hampered by the complex geodynamic setting of the Mediterranean 
region \citep[see][]{anzidei2014coastal,faccenna2014mantle}, where tectonics and isostasy are contributing simultaneously to 
vertical deformations and gravity variations, thus producing a complex pattern of relative \sealevel change. 
With the aim of separating these effects, long-term relative \sealevel data from the Mediterranean Sea have been often 
interpreted with the aid of global Glacial Isostatic Adjustment (GIA) models. These are based upon the Sea Level Equation (SLE) 
first introduced by~\citet{farrell76on}, which is 
solved adopting a specific deglaciation chronology for the late-Pleistocene ice sheets and an \textit{a priori} rheological profile 
\citep{spada2017glacial,whitehouse2018}. 

Due to the delayed viscoelastic response of the Earth's system to surface mass redistributions, signals from the last deglaciation 
are still {detectable} today across the Mediterranean region. Previous model computations of \citeauthor{stocchi2007glacio} 
(\citeyear{stocchi2007glacio}, \citeyear{stocchi2009influence}) have shown that these GIA imprints can significantly affect 
\sealevel measurements at tide gauges, vertical and horizontal land motion observed by means of GNSS methods and absolute 
\sealevel variations tracked by altimeters~\citep{cazenave2002sea}.  Although the importance of the ongoing isostatic readjustment has been 
clearly recognized in a number of works \citep[see \emph{e.g.,}][]{serpelloni2013vertical,anzidei2014coastal}, 
these contemporary regional effects of GIA have received comparatively little attention so far. Our new assessment, which 
builds upon previous work of \citet{stocchi2009influence}, is motivated by the increasing number of high quality space-borne 
and ground-based geodetic data available across the Mediterranean region~\citep{tsimplis2013effect,bonaduce2016sea},  
the recent development of {new global models} {of the ice history and the layered Earth structure parameters} ~\citep{peltier2015space,royandpeltierice75} and the availability of new numerical tools 
\citep{spada2019selen4-gmd-12-5055-2019}. 

GIA models account for gravitational, deformational and rotational interactions within the Earth system and explain 
the spatial and temporal variability of sea level in response to surface mass redistributions {(see \citet{spada2017glacial} and \citet{whitehouse2018} for a review)}. 
In global studies, the fine details of the GIA imprint across the Mediterranean are scarcely appreciated, due to the relatively small 
extent of the basin \citep[see \emph{e.g.,}][]{tamisiea2011ongoing}. In this region, the ongoing \sealevel variations due to 
GIA have been first visualized -- but not discussed -- in the work of \citet{mitrovica2002origin}, who solved globally the SLE to a 
sufficiently high spatial resolution. Based upon a modified version of model ICE-3G~(VM1) of~\citet{tushingham1991ice}, 
the global maps of {Mitrovica and Milne}
clearly show that GIA is 
still causing significant \sealevel variations across the Mediterranean Sea. Maximum amplitudes of relative \sealevel rise are reached 
in the bulk of the basin and decline toward the coastlines. The same peculiar pattern is predicted at adjacent mid-latitude basins, \emph{i.e.,} 
the Black Sea and the Caspian Sea. The GIA-induced relative \sealevel rise originates from a basin-scale land subsidence, with the maximum 
rates attained in the bulk of the basin. These patterns of relative \sealevel rise and land subsidence {have been interpreted by \citet{stocchi2007glacio} as an 
effect of the extra-loading exerted by meltwater on the seafloor during deglaciation {(hydro-isostasy})}, which is still {ongoing due to the 
{persisting} non-isostatic conditions}. {A role of}
the melting of the former Fennoscandian ice sheet has been invoked {by \citet {stocchi2007glacio}}, while the contribution from 
the deglaciation of the nearby Alpine ice sheet still remains uncertain~\citep{stocchi2005isostatic,sternai2019present}
despite the improvements in glaciological modelling \citep[see][and references therein]{seguinot2018modelling}. 

The present regional imprint of GIA across the Mediterranean Sea has been studied in detail by \citet{stocchi2009influence},
adopting some of the first ICE-$X$ models developed by {WR Peltier and collaborators}. They have 
considered predictions of a suite of GIA models in terms of rate of relative and absolute \sealevel change and vertical land motion, both 
at a basin scale and at tide gauges locations, but paying no attention to possible horizontal motions. 
Since then, however, a number of improved GIA models consistent with 
global relative \sealevel datasets {has} been introduced, 
including revised deglaciation chronologies and viscosity profiles, which until now have not been fully exploited to study the ongoing 
effects of GIA in the region~\citep{peltier2015space,royandpeltierice75}. Furthermore, the former GIA simulations of \citeauthor{stocchi2007glacio} (\citeyear{stocchi2007glacio}, 
\citeyear{stocchi2009influence}) were obtained adopting a coarse spatial resolution, and {some potentially important} effects such as the 
rotational feedback on sea level and the horizontal migration of the shorelines were {not considered}. 
Of course, we should not expect that taking these features into account would profoundly affect our knowledge about the effects of GIA across
the Mediterranean Sea. However, 
improving the modeling scheme by introducing an ensemble approach
would certainly provide more robust results. Furthermore, upgraded GIA computations  
would facilitate the interpretation of geodetic data and provide up-to-date corrections to a number of 
observations. Last, the results of~\citet{melini2019someremarks} suggest that an ensemble approach would be useful to constrain 
the uncertainties that are still involved in GIA modelling~and to explore the future trends of sea level expected in the region. 
{The importance of regional GIA modeling uncertainties has been also 
discussed by \citet{love2016contribution}, \citet{vestol2019nkg2016lu}, \citet{simon2020uncertainty} and \citet{kierulf2021gnss}, although these works were not focused on the Mediterranean Sea.}

The paper is organized as follows. The methods are briefly described in Section~\ref{sec:methods}. In Section~\ref{sec:patterns} 
we describe the imprints of GIA on several geophysical quantities in the Mediterranean region. Ensemble GIA modelling results 
are presented in Section \ref{sec:ensemble} and discussed in Section~\ref{sec:discussion}. Our conclusions are drawn 
in Section~\ref{sec:conclusions}.  

\section{Methods}\label{sec:methods}

All the GIA simulations in this work have been performed using the open source SLE solver 
SELEN$^4$~\citep[SELEN version 4, see][]{spada2019selen4-gmd-12-5055-2019}, in which 
the published histories of deglaciation of the various GIA models and the rheological profiles of 
the mantle have been incorporated. However, in some instances, we have
used GIA results directly available from the Datasets page of WR 
Peltier\footnote{See~\emph{https://www.atmosp.physics.utoronto.ca/$\sim$peltier/data.php}, last accessed on April 20, 2021.},
which also provides details about the setup of the most recent models of the ICE-$X$ suite.  

SELEN$^4$ solves the gravitationally and topographically self-consistent SLE, which in its simplest form reads 
\begin{equation}\label{eq:s}
S(\theta,\lambda,t) = N-U\,,
\end{equation}
where $S(\theta,\lambda,t)$ is relative \sealevel change at the location of colatitude $\theta$ and longitude $\lambda$, $t$ is time, {
while $N=N(\theta,\lambda,t)$ and $U=U(\theta,\lambda,t)$ are absolute} \sealevel change and {the} vertical displacement of the Earth's surface, respectively.
A thorough discussion of the physics of SLE is given {in the seminal work of \citet{farrell76on} and in \citet{tamisiea2011ongoing}}.  

Since both $N$ and $U$ implicitly depend on $S$, the SLE (\ref{eq:s}) is an integral equation that is solved numerically through an iterative scheme, which demands a suitable spatiotemporal discretization of the involved fields.
We {have defined} a time discretization assuming constant steps of length $\Delta t=500$ years, over an integration interval that extends back to the Last Glacial Maximum. 
For the space discretization, we have taken advantage of the equal-area, icosahedron-based geodetic grid of 
\citet{tegmark1996icosahedron}, using a resolution parameter $R=100$ that corresponds to cells of size $\sim 20$ 
km on the surface of the Earth. The maximum degree of the analysis has been set to $l_{max}=512$, corresponding to 
a spatial {wavelength} of $\approx 75$ km. In some runs, to alleviate
the computational burden, the parameters $R=60$ and $l_{max}=256$ have been adopted, without significant loss of 
precision in the final results. 
To prescribe the ``final'' (\emph{i.e.}, present-day) condition of Earth's topography, we have adopted the bedrock version of the global 
ETOPO1 dataset \citep{amante2009etopo1,eakins2012hypsographic}, integrated with the Bedmap2 topographic model 
\citep{fretwell2013bedmap2} {south of $60^\circ$S}.
The rotational feedback on \sealevel change has been modeled according to {the revised rotation theory
of} \citet{mitrovica2005rotational} and \citet{mitrovica2011ice}. 
The solution algorithm of the SLE consists of two nested iterations, where the external one updates the paleo-topography  according to the 
solution of the SLE, which is performed in the internal iteration. In all the runs, we have adopted five external and five
internal iterations to ensure convergence. Once $S$, $N$ and $U$ are obtained by solving iteratively the SLE (Eq.~\ref{eq:s}), 
a suite of additional geodetic quantities associated with GIA are accessible, as the East ($U_e$) and North ($U_n$) components 
of the horizontal displacement field and their present trends. {For} more details about the solution method, the reader is referred
to the supplementary material of \citet{spada2019selen4-gmd-12-5055-2019}. 

\section{GIA patterns in the Mediterranean region}\label{sec:patterns} 

\begin{figure}
\centering
\includegraphics[width=0.8\textwidth,angle=0]{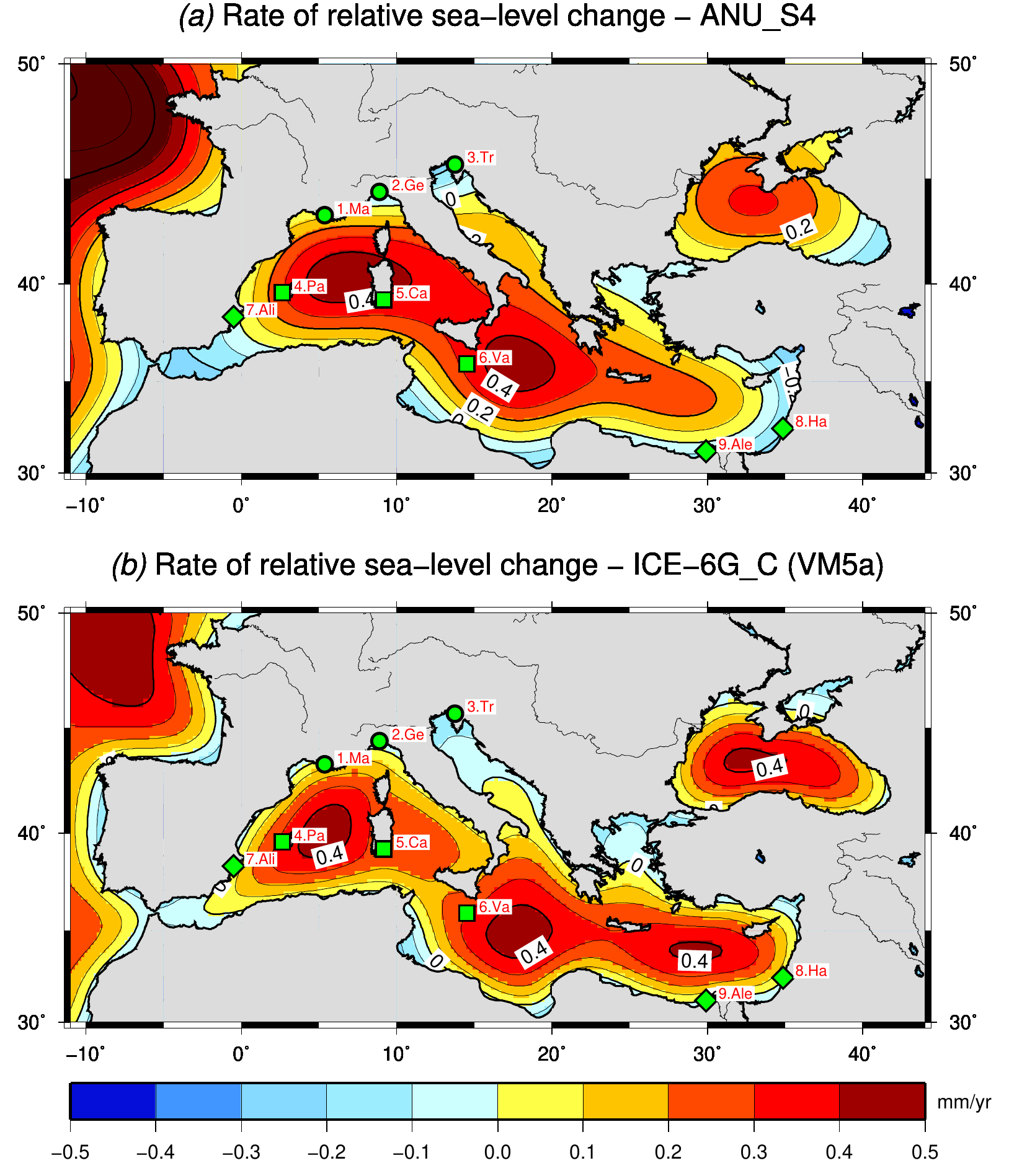}
\caption{Patterns of the rate of present day \sealevel change $\dot S$ induced  by GIA across the Mediterranean region,
according to models ANU$\_$S4 (a) and ICE-6G$\_$C~(VM5a) (b). The data in {(a)}  
have been obtained from the Datasets page of {WR Peltier}. In both frames, the locations of a few tide gauges deployed along the coastlines are marked by green symbols. {The station names are abbreviated (full names are given in the text and in Table \ref{tab:rates}).}}
\label{figure:peltier+lambeck}
\end{figure}

\begin{figure}
\centering
\includegraphics[width=0.9\textwidth,angle=0]{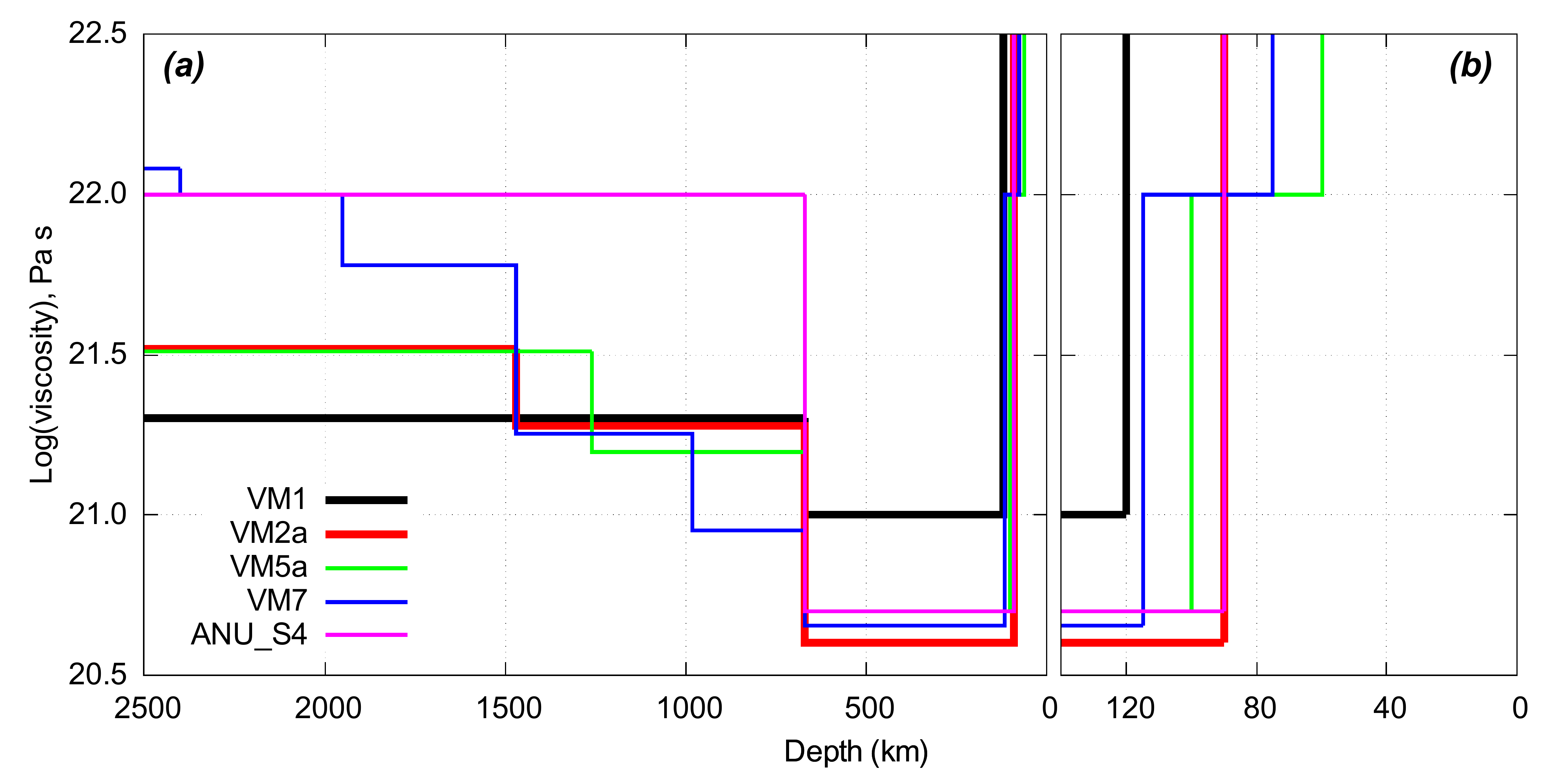}
\caption{{Radial viscosity profiles considered in this study. Frame (a) shows the full structure while (b) shows an enlarged view of the depth range $0$--$140$ km.  VM1 assumes lower and upper mantle viscosities of $2 \times 10^{21}$ Pa$\cdot$s and $10^{21}$ Pa$\cdot$s, respectively \citep{tushingham1991ice}.
VM2a is a simplified version of the multi-layered VM2\_L90 profile available from the web page of WR Peltier.
Numerical values for VM5a and VM7 are from Table 2 of \citet{royandpeltierice75}. ANU\_S4 assumes lower and upper mantle viscosities of $10^{22}$ and $5\times 10^{21}$ Pa$\cdot$s, respectively. Profiles VM5a and VM7 include a $40$ km-thick high-viscosity layer ($10^{22}$ Pa$\cdot$s) at the base of the lithosphere.}}
\label{figure:viscosity-profiles}
\end{figure}

Figure~\ref{figure:peltier+lambeck}~shows predictions for the ongoing rate of relative \sealevel change across the Mediterranean Sea, hereafter denoted by ${\dot S}$, according to two state-of-the-art GIA models. The first model (Figure~\ref{figure:peltier+lambeck}a), is the one progressively developed at the Australian National University (ANU) by {Kurt Lambeck} and collaborators \citep[see \emph{e.g.,}][]{nakada1987glacial,lambeck2003water}. This map has been obtained by implementing the ANU ice chronology, provided to us by Anthony Purcell in November $2016$, into SELEN$^4$; hence, this model shall be referred to as ANU$\_$S4 in the following.{
The setup of ANU$\_$S4 is based on a realization of the spatio-temporal evolution of ice complexes on a icosahedron-based global grid and assuming a piecewise constant time history, as described in Section \ref{sec:methods} \citep[see also][for further details]{melini2019someremarks}. The radial viscosity profile used in association with ANU\_S4, shown in Figure~\ref{figure:viscosity-profiles}, assumes a $90$ km elastic lithosphere and a viscosity of $5\times 10^{20}$ Pa$\cdot$s and $10^{22}$ Pa$\cdot$s in the upper and lower mantle, respectively \citep{lambeck2017glacial}.~
}
The second model (Figure~\ref{figure:peltier+lambeck}b) is ICE-6G$\_$C~(VM5a), where ``ICE-6G$\_$C'' and ``VM5a'' denote {its two} basic components, namely the deglaciation chronology and the {layered Earth structure parameters}, respectively. This model, described by~\citet{argus2014antarctica} and \citet{peltier2015space} is one of the latest iterations of the suite of ICE-$X$ models historically developed at the University of Toronto by Prof. 
{WR Peltier} and collaborators. 

The general patterns of \sealevel change shown in Figure~\ref{figure:peltier+lambeck} confirm the results of~\citet{mitrovica2002origin} and \citet{stocchi2009influence}, clearly indicating that GIA is currently responsible for a general relative \sealevel rise across the Mediterranean Sea. However, due to the high spatial resolution of these maps, the details of the non-uniform pattern of \sealevel change caused by hydro-isostasy can be better discerned. The maximum values of ${\dot S}$, slightly exceeding the value of $0.4$ \rates, are attained across the widest sub-basins of the Mediterranean Sea, \emph{i.e.,} the Balearic, the Ionian, the Levantine and the Black seas. \citet{stocchi2007glacio} have interpreted this pattern as the still progressing 
lithospheric flexure induced by the meltwater load, causing a \sealevel rise relative to the seafloor, with maximum effects in the heart of the basins 
and its amplitude that declines approaching the shorelines. Of course, since the continental masses are unevenly distributed, the flexure due to subsidence 
shows a complex geometry, which is also partly determined by gravitational and rotational effects implicit in the SLE. It should be recalled that the two GIA models considered here are based upon different eustatic (\emph{i.e.,} ice-volume equivalent) curves, different deglaciation chronologies and rely upon distinct assumptions regarding the rheological profiles for the mantle. Despite 
these and other structural distinctions~\citep[for details, see][]{melini2019someremarks}, in the Mediterranean region the $\dot S$ patterns are  found to be broadly comparable. However, significant differences can be noted across the Levantine Sea and the Black Sea, where predictions based upon 
ICE-6G$\_$C~(VM5a) generally exceed those obtained using ANU$\_$S4. 

A noticeable feature of the GIA imprints in Figure~\ref{figure:peltier+lambeck} is the relatively small or even negligible value of ${\dot S}$ 
generally attained along the continental coastlines of Southern Europe and of North Africa, compared to the open sea. Interestingly, swathes 
of relative \sealevel \emph{fall} can be generally noted in places where the coastlines are characterized by a relatively 
short radius of curvature. For example, this is 
observed along the coasts of the Alboran Sea and {of} Tunisia, but also in the northern Aegean Sea and in the northern Adriatic Sea. 
In these marginal and narrow seas, the current trend of relative \sealevel driven by GIA is dominated by the effect continental levering, which is manifested by subsidence 
of offshore locations and the upward tilting of onshore locations \citep[see][]{WALCOTT1972,mitrovica1991postglacial,mitrovica2002origin,murray2014quaternary,CLEMENT_2016_73}. Condition ${\dot S}<0$ suggests 
that, at some time during the late Holocene, sea level may have been higher than today at these locations. \citet{kearney2001late} 
has discussed about the possible existence of \sealevel high-stands in the northern hemisphere during the Holocene and 
some works have reported evidence of high-stands, although without invoking isostatic mechanisms 
\citep[see \emph{e.g.,}][]{pirazzoli1991holocene,1993-bernier-367165,1997-sanlaville-etal,morhange2006late}.  
Remarkably, despite the structural differences, 
ICE-6G$\_$C~(VM5a) and ANU$\_$S4 also broadly agree upon the position of possible Holocene high-stands, which are identified by the condition $\dot S<0$. In these locations, GIA is counteracting the general \sealevel rise caused by the present-day 
terrestrial ice melt, which is characterized by very smooth imprints across the Mediterranean Sea~\citep{galassi2014sea}. 
Although tectonic deformations or factors associated to the ocean circulation  
could potentially overprint the isostatic contribution to sea level, evidence from specific sites like the Gulf of Gab\`es along the SE coast of Tunisia 
effectively confirms the existence of a late-Holocene high-stand~\citep[see \emph{e.g.}][]{mauz2015terminal}. The existence of other possible high-stands
along the coasts of the Mediterranean Sea, suggested by the map of Figure~\ref{figure:peltier+lambeck}, shall be the topic of a follow-up study. 

\begin{figure}
\centering
\includegraphics[width=0.9\textwidth,angle=0]{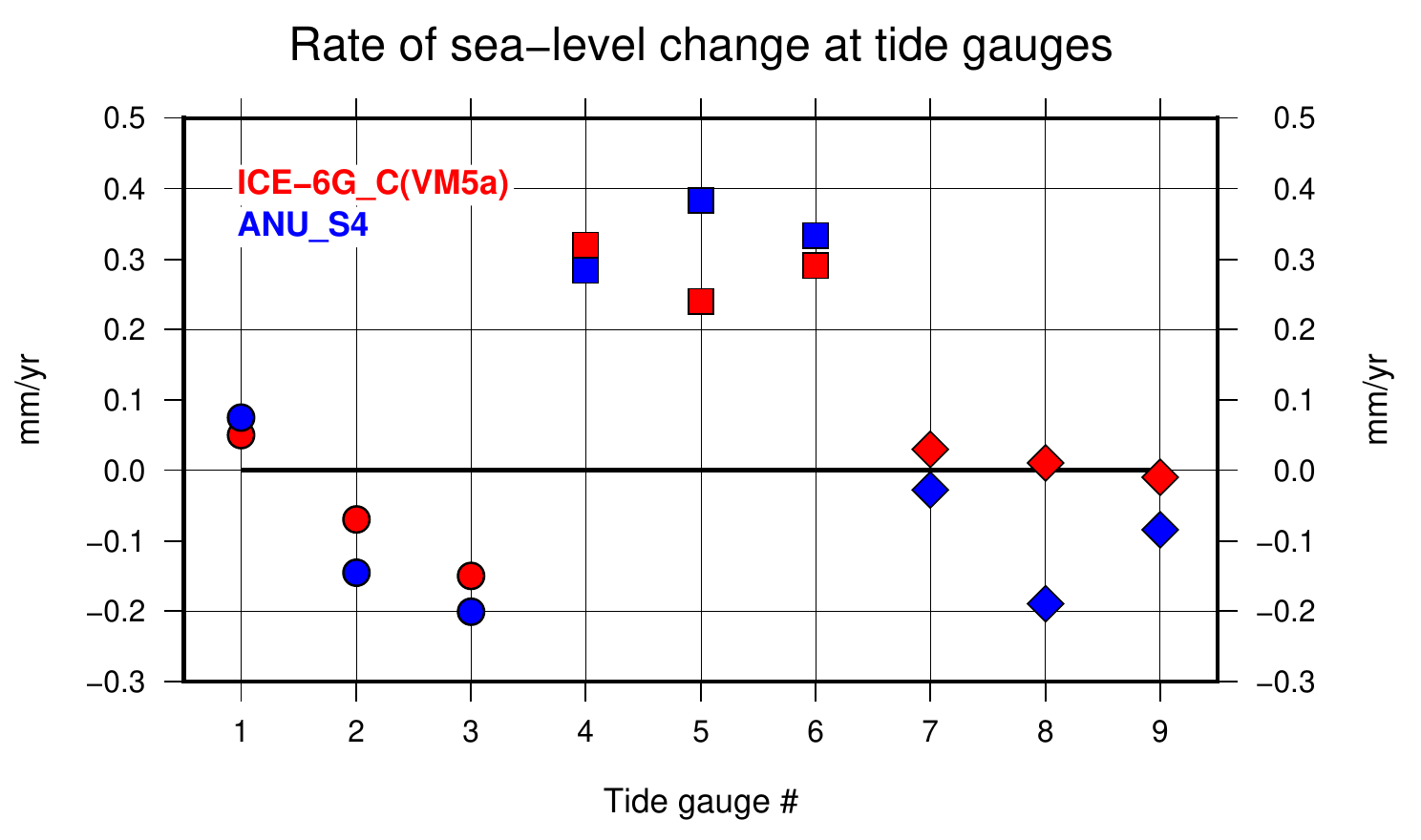}
\caption{Rates of relative \sealevel change $\dot S$ expected at tide gauges according to models ICE-6G$\_$C~(VM5a) (red) and 
ANU$\_$S4 (blue). {The tide gauges locations are marked by green symbols in Figure~\ref{figure:peltier+lambeck}, which also shows the station names abbreviations. Numerical values of the rates are listed in Table \ref{tab:rates}.}}
\label{figure:peltier+lambeck+tg}
\end{figure}

A more quantitative ~inter-comparison ~between the ~predictions ~of models ~ICE-6G$\_$C~(VM5a) ~and~ANU$\_$S4 is 
drawn in~Figure~\ref{figure:peltier+lambeck+tg}, showing the values of ${\dot S}$ at 
specific locations where several tide gauges are sited (see green symbols in Figure~\ref{figure:peltier+lambeck}). 
The sites of Marseilles (1), {Genova} (2) and Trieste (3), marked by circles, have a particular importance since they are all characterized 
by long records and therefore they have been considered in various estimates of secular global mean~\sealevel rise \citep[see~\emph{e.g.,}][]{douglas1991global,woodworth2003some,Douglas_1997,Spada_Galassi_2012}. Despite the short distance separating the sites (see Figure~\ref{figure:peltier+lambeck}), 
the contribution of GIA to ${\dot S}$ is not uniform at these locations. However, it has a relatively modest amplitude, varying in the range between $\sim -0.2$ and $\sim +0.1$ \rates, 
according to both GIA models. In the western (Alicante I, 7) and in the eastern Mediterranean, at Hadera (8) and Alexandria (9) (diamonds), 
a similar range of responses is found for ANU$\_$S4 while ICE-6G$\_$C~(VM5a) points to negligible values. 
It is apparent that consistent with the pattern of Figure~\ref{figure:peltier+lambeck}, {significant} values of ${\dot S}$ are only expected 
at tide gauges located in the bulk of the basin. Indeed, at the sites of Palma de Mallorca (4), Cagliari (5) and Valletta (6) (squares), 
rates as large as ${\dot S}\sim 0.4$ \rate are predicted by both models. For Cagliari, this is a significant fraction of 
the long-term rate of \sealevel rise effectively observed~\emph{in situ}\footnote{The trends 
of relative \sealevel change of all the PSMSL stations and their standard errors are available from 
~\emph{https://www.psmsl.org/products/trends/trends.txt} (last accessed on June 3, 2021).} ($1.88 \pm 0.24$ \rates), where we note that 
the GIA effect exceeds the standard error of the trend. A similar rate  ($1.57 \pm 1.12$~\rates) is observed 
at Valletta (or Marsaxlokk), although the GIA contribution is smaller than the uncertainty. 
For Mallorca, the time span of the data ($1997$-$2018$) is too short to establish a reliable trend.
{For reference, the numerical values of the ${\dot S}$ values portrayed in Figure~\ref{figure:peltier+lambeck+tg} are reported in Table~\ref{tab:rates} below, along with the 
values obtained by the ensemble modelling approach described in Section \ref{sec:ensemble}.}

\begin{figure}
\centering
\includegraphics[width=1\textwidth,angle=0]{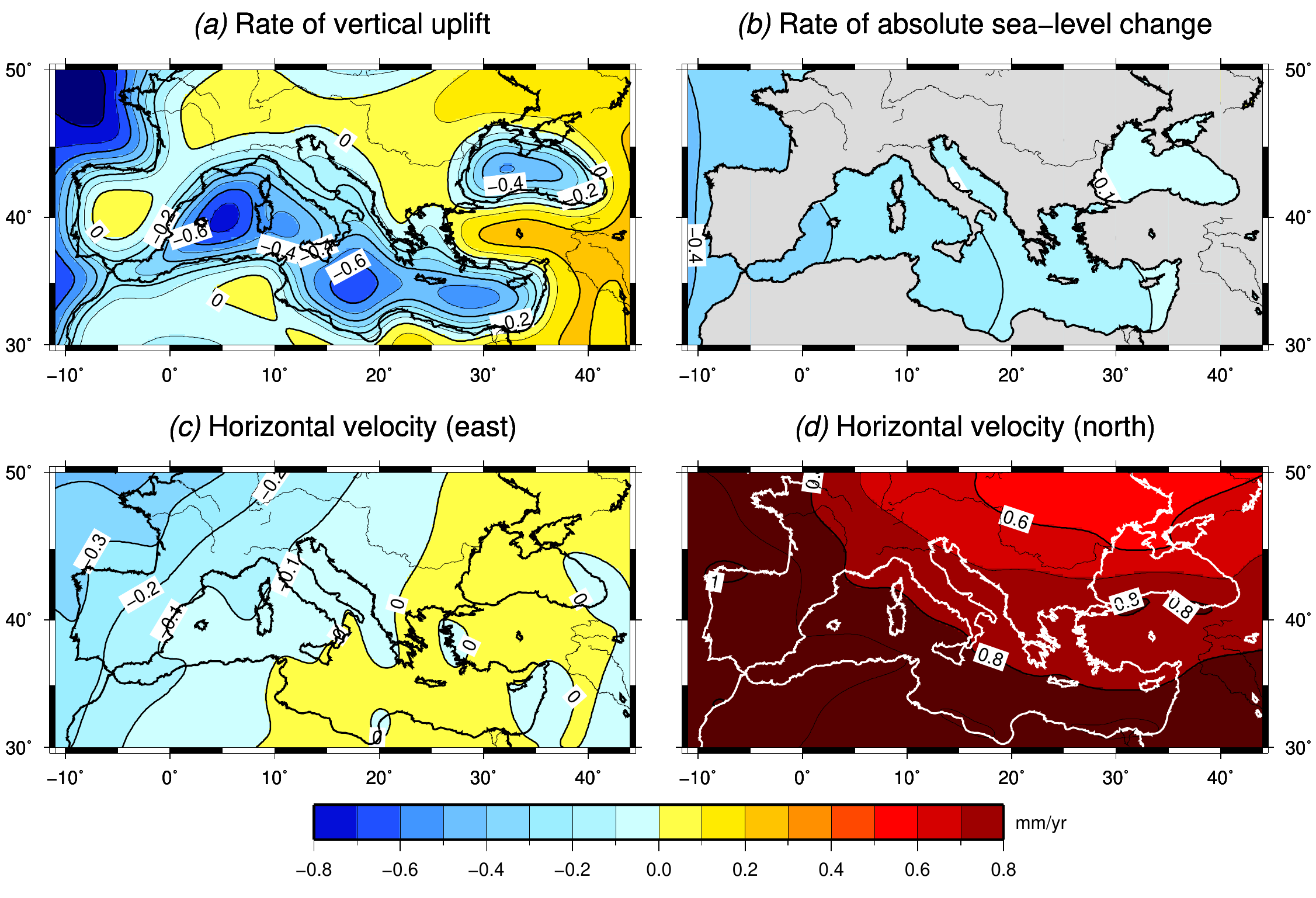}
\caption{Predictions for the rate of vertical displacement ($\dot U$, a),  of absolute \sealevel change  ($\dot N$, b),  
of the east component of horizontal displacement ($\dot U_e$, c) and for the north component ($\dot U_n$, d). All the maps are based upon model 
ICE-6G$\_$C~(VM5a). Data are from the Datasets page of {WR Peltier}.}
\label{figure:peltier+more}
\end{figure}

To better characterize the ongoing geodetic variations in the Mediterranean Sea, 
in~Figure~\ref{figure:peltier+more} we now consider a further set of variables 
associated with GIA, in addition to ${\dot S}$. These are the rate of vertical uplift ($\dot U$), the rate of absolute \sealevel change
($\dot N$), as well as the East ($\dot U_e$) and North ($\dot U_n$) components of the horizontal rate of displacement. The latter {components} were not 
considered in~\citet{stocchi2009influence} nor in subsequent works regarding the Mediterranean region. 
We note that ${\dot S}$, ${\dot N}$ and ${\dot U}$ are not independent of 
one another, being connected through the Sea Level Equation (\ref{eq:s}). 
The complexity of the four patterns in Figure~\ref{figure:peltier+more}, all pertaining to model 
ICE-6G$\_$C~(VM5a), is apparent, and confirms that GIA has an important role in the spatial variability of present-day geodetic signals across the 
Mediterranean Sea. Similar results, not shown here, are obtained for model ANU$\_$S4. 
It is apparent that $\dot U$ (Figure~\ref{figure:peltier+more}a) is strongly anti-correlated with $\dot S$ (see~Figure~\ref{figure:peltier+lambeck}a), 
showing a widespread (but not uniform) state of subsidence across the whole Mediterranean basin, with maximum values of $-0.6$ \rates. We note 
that with the exception of some very narrow inlets, GIA is causing a general subsidence of all the coastlines ($\dot U < 0$), including those stretches 
where $\dot S < 0$, \emph{i.e.,} where a Holocene high-stand is expected (see Figure~\ref{figure:peltier+lambeck}a). The pattern of $\dot N$ (b) shows that 
absolute sea level is falling across the whole Mediterranean basin, opposite to relative \sealevel change $\dot S$. 
Furthermore, in contrast with $\dot U$, the rate of absolute \sealevel change $\dot N$ shows little spatial variability and attains relatively small values
with respect to $\dot{U}$. These are comparable with the global ocean-average $<\dot N>~ \approx~ -0.3$ \rate \citep{melini2019someremarks,spada2019some}, which 
constitutes the GIA correction to absolute \sealevel variations detected by satellite altimetry \citep[see~\emph{e.g.,}][]{tamisiea2011ongoing}. According to the patterns in Figures~\ref{figure:peltier+more}c and \ref{figure:peltier+more}d, the rates of horizontal displacements have a significant 
amplitude. In particular, the neatly positive $\dot U_n$ values (d) indicate that GIA is currently imposing a nearly uniform northward drift 
across the whole Mediterranean region, at
a rate of $\sim~0.8$ \rates. By a global analysis, we have confirmed that this geodetically significant drift, {whose amplitude remarkably exceeds} $\dot U$ and $\dot S$, is to be attributed to the effect of the melting of the Laurentian and northern Europe ice sheets. The east component $\dot U_e$ 
has a minor role, with an amplitude not exceeding $\sim 0.1$ \rate across the Mediterranean Sea (c).
The GIA-induced northward
drift notably exceeds the vertical rates. However, it is not
expected to affect significantly the velocity fields observed by GNSS
networks, which are dominated by larger tectonic signals \citep[see, \textit{e.g.}][]{faccenna2014mantle}.

\begin{figure}
\centering
\includegraphics[width=1\textwidth,angle=0]{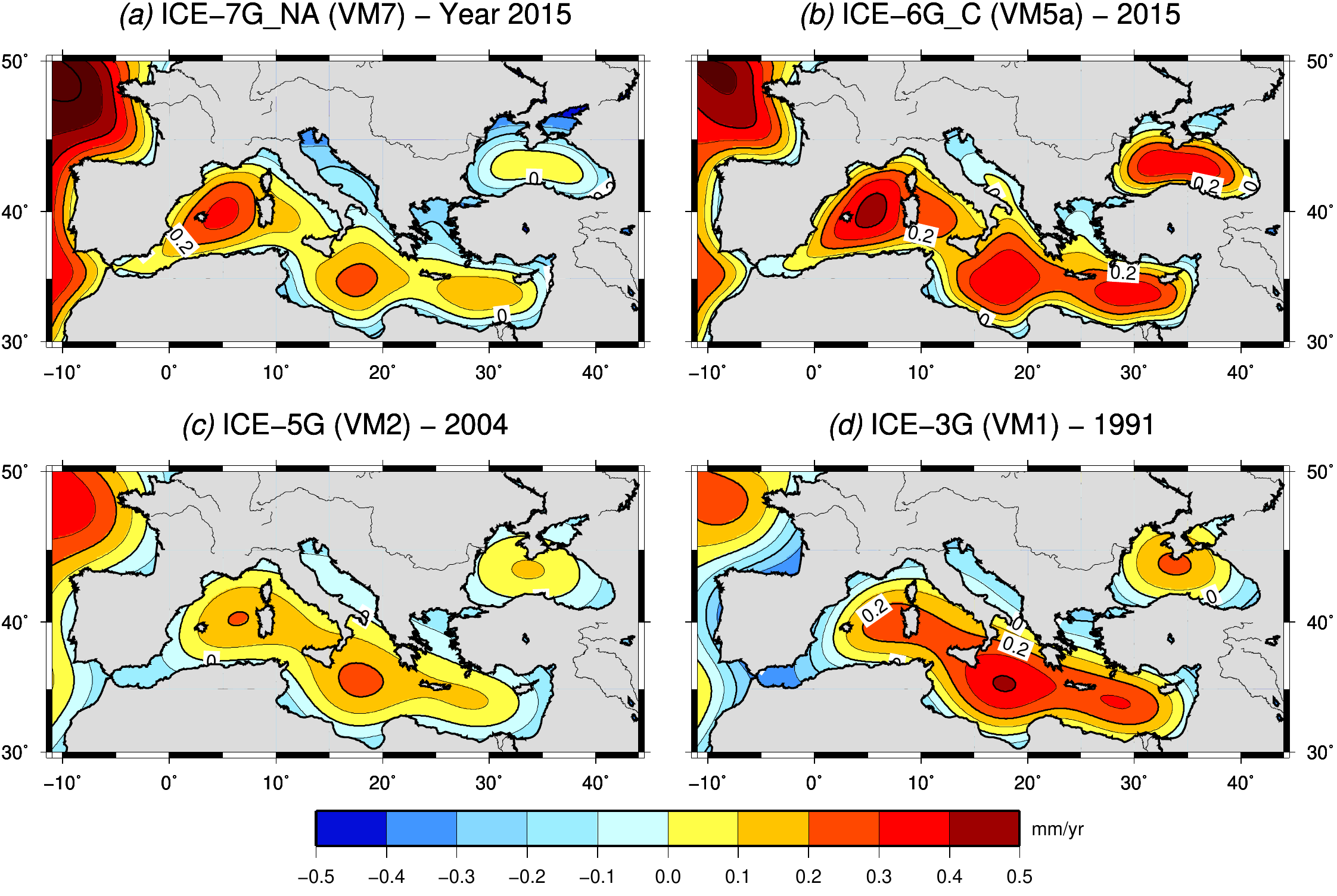}
\caption{Rates of present day relative \sealevel change $\dot S$ in the Mediterranean region, according to some combinations of  
ice deglaciation histories and viscosity profiles belonging to the ICE-$X$ suite
of WR Peltier and collaborators. The model name and the date of publication are given in the headers. All the computations have been performed using program SELEN$^4$.}
\label{figure:peltier+icex}
\end{figure}

Since GIA models are continuously evolving, their predictions are not given once and for all~\citep{melini2019someremarks}. 
To get a flavor of how the evolution of GIA models has influenced the pattern of ${\dot S}$ across the Mediterranean region,  
in Figure~\ref{figure:peltier+icex} we consider results for some members of the suite
of ICE-$X$ models historically developed by {WR Peltier} and collaborators. They include model ICE-7G$\_$NA (VM7) 
of \citet{royandpeltier2015} and its precursors ICE-6G$\_$C~(VM5a) \citep{peltier2015space}, ICE-5G (VM2)
\citep{peltier2004global} and ICE-3G (VM1) \citep{tushingham1991ice,tushingham1992validation}. 
For results 
based upon the first model of the suite (\emph{i.e.,} ICE-$1$ of \citeauthor{peltier1976glacial}~\citeyear{peltier1976glacial}), see \citet{stocchi2009influence}. These GIA models 
have been introduced to progressively improve the fit with global sets of Holocene \sealevel proxies and geodetic data. They are
characterized by distinct rheological profiles and different melting histories of the continental
ice sheets (see the references quoted above for details). All the runs in Figure~\ref{figure:peltier+icex} 
have been performed using the open source program SELEN$^4$~\citep{spada2019selen4-gmd-12-5055-2019}, 
in which the published histories of deglaciation and the rheological profiles of the mantle, 
{shown in Figure~\ref{figure:viscosity-profiles},}
have been assimilated. 
We remark that differences between the SELEN$^4$ 
results and those published in the original works are possible, as it can be seen comparing Figure~\ref{figure:peltier+icex}b with 
Figure~\ref{figure:peltier+lambeck}a, both pertaining to ICE-6G$\_$C~(VM5a). These may reflect differences in the numerical schemes adopted
to solve the SLE, in the theory employed to describe the rotational feedback on \sealevel change, 
in the geometry and resolution of the grid on which the SLE is discretized, and in the nature of mantle layering \citep[see][]{melini2019someremarks}. 
These differences shall be understood as soon as a suite of benchmark computations will be established 
among SLE solvers, along the lines of previous efforts within the GIA community \citep{spada2011benchmark,martinec2018benchmark,Kachuck19benchmarked}. 
From 
Figure~\ref{figure:peltier+icex} it is apparent that all the ICE-$X$ models broadly agree on the ${\dot S}$ patterns, which are all 
characterized by a \sealevel rise in the bulk of the basin, also suggesting possible Holocene high-stands in the narrower 
inlets. However, it is clear that the patterns vary significantly in the details and that peak values attained in the various 
sub-basins of the Mediterranean Sea also differ. Overall, the variance of the results in 
Figure~\ref{figure:peltier+icex} clearly indicates that GIA predictions are affected by a significant degree
of uncertainty on the Mediterranean scale. This justifies an ensemble approach
involving a larger population of state-of-the-art GIA models.

\section{Ensemble GIA modelling in the Mediterranean region}\label{sec:ensemble} 

During the last decade, the importance of evaluating the uncertainties associated with GIA modelling has been 
recognized in a number of studies, and assessed through ensemble-like approaches. Until now, this has been done in various 
regional and global contexts, but never in the Mediterranean Sea. For example, in a re-analysis of Gravity Recovery and Climate 
Experiment (GRACE) measurements, \citet{sasgen2012antarctic} have inverted the mass balance of 
the Antarctic ice sheet during 2002--2011 solving the forward GIA problem for a very rich set of models.   
Subsequently, using Bayesian methods and testing a large amount of GIA models, \citet{caron2018gia} have evaluated 
uncertainties associated with imperfect knowledge of mantle viscosity upon the Stokes coefficients of 
the Earth's gravity field. Uncertainties in 1-D GIA modelling have also been evaluated in an ensemble 
modelling perspective by \citet{melini2019someremarks}, with the purpose of 
assessing their influence upon estimates of secular \sealevel rise. Shortly after, \citet{li2020uncertainties} 
considered GIA uncertainties in the regional context of North America, accounting for the 
possible effects of 3-D Earth's structure.  Using an ensemble approach, \citet{esd-11-129-2020} have built a global 
semi-empirical GIA model based on GRACE data. \citet{melini2019someremarks} and 
\citet{li2020uncertainties} have classified the GIA modelling uncertainties into two types. The first type (T1)
is associated with the input parameters of the GIA models, \emph{e.g.,} the Earth viscosity profile or the 
loading history of continental ice sheets. The second type (T2) is associated with structural differences 
among GIA models. These include, for example, different numerical approaches to the 
solution of the SLE, the use of different eustatic curves, the adoption of different sets of 
geophysical constraints, or diverging \emph{a priori} assumptions about the Earth’s 
viscosity profile. Possible ambiguities in the proposed classification of GIA modelling 
uncertainties however exist, as discussed by~\citet{melini2019someremarks}. 

Characterizing GIA uncertainties at the Mediterranean scale is a challenging task. Indeed, 
a major issue arises because of the imperfect knowledge about regional-scale rheological 
heterogeneities, which certainly exist in such a complex tectonic setting~\citep[see][and references therein]{faccenna2014mantle}. 
This would motivate, in the study area, a fully 3-D GIA modelling approach, which is however far
from being realized. Limiting our attention to 1-D GIA modelling, here we take inspiration from previous 
multi-modelling approaches by \citet{lambeck-purcell-2005} and \citet{royandpeltier2015}, who have
considered GIA modelling uncertainties into two different contexts. The study of 
\citet{lambeck-purcell-2005} is of particular relevance here, since it investigates the sensitivity 
of Holocene sea level at some Mediterranean sites to variations of the GIA model parameters. 
\citeauthor{lambeck-purcell-2005} found that, at specific locations along the Mediterranean coastlines,  
relative \sealevel predictions based on distinct rheological profiles 
can vary up to a few meters between $12$ and $6$ ka. However, their work was not {dedicated} 
 to the evaluation to the GIA 
contribution to \sealevel trends at present time, which is the target of our analysis. More recently, \citet{royandpeltier2015} 
computed synthetic relative \sealevel curves at selected North American sites using a suite of variants of the VM5a
viscosity profile. They found that the thickness of the lithosphere and the viscosity structure of the upper 
mantle have an important influence on the relative sea level predictions, while the viscosity 
at depth significantly affects the spatiotemporal evolution of the lateral fore-bulge. It is of interest 
here to test whether these variants of the VM5a rheological profile can also influence the 
present-day \sealevel change predictions in the Mediterranean region, away from the former centers of 
deglaciation. 

To model the GIA contribution to \sealevel rise in the Mediterranean region and its uncertainty, we have built 
two independent ensembles based upon previous works of \citet{lambeck-purcell-2005} 
and~\citet{royandpeltier2015}, respectively. Since these works have been carried out independently and 
have proposed structurally distinct GIA models, based on different datasets, merging them in a unique
ensemble would not be appropriate. Results obtained from the two ensembles should not be expected to overlap
and could show a different sensitivity to the parameters that define each of the models. 
The first ensemble (E1) encompasses different realizations of model 
ANU$\_$S4, all sharing the same deglaciation chronology. Following \citet{lambeck-purcell-2005}, 
the thickness of the elastic lithosphere and the upper and lower mantle viscosities are varied within 
the ranges listed in Table~\ref{tab:ens}. 
{Coherently with \citet{lambeck-purcell-2005}, in building ensemble E1 we assume a nominal model with an upper mantle viscosity of $3\times 10^{20}$ Pa$\cdot$s and a lithospheric thickness of $65$ km. These values differ from those that we have used in ANU\_S4, which is based on the rheological profile used for global-scale GIA models by the ANU group \citep[see \textit{e.g.}][]{lambeck2017glacial}. We shall therefore refer to this variant of of ANU\_S4 as ANU\_S4(E1).}
{Ensemble E1 consists of $42$ 
GIA models.} 
The second ensemble (E2), which consists of $74$ models, is originated from
ICE-6G$\_$C~(VM5a), considering variations of the lithospheric thickness and 
of the viscosity of each of the four mantle layers that characterize VM5a. The range of variability of
each parameter is chosen to encompass the discrete values explored by \citet{royandpeltier2015} 
(see Table~\ref{tab:ens}), keeping the deglaciation chronology of ICE-6G$\_$C unaltered. 
In both ensembles, we vary a single rheological parameter 
within a pre-assigned range while keeping all the others fixed to their nominal value. When the 
lithospheric thickness is varied, we keep its elastic constants fixed to the PREM (Preliminary 
Reference Earth Model, see \citeauthor{dziewonski1981preliminary}~\citeyear{dziewonski1981preliminary})  
averages obtained for the nominal model
{and, for ensemble E2, we retain a $40$ km-thick, high-viscosity layer at the base of the lithosphere (see Figure~\ref{figure:viscosity-profiles}).}

\begin{figure}
\centering
\includegraphics[width=1\textwidth,angle=0]{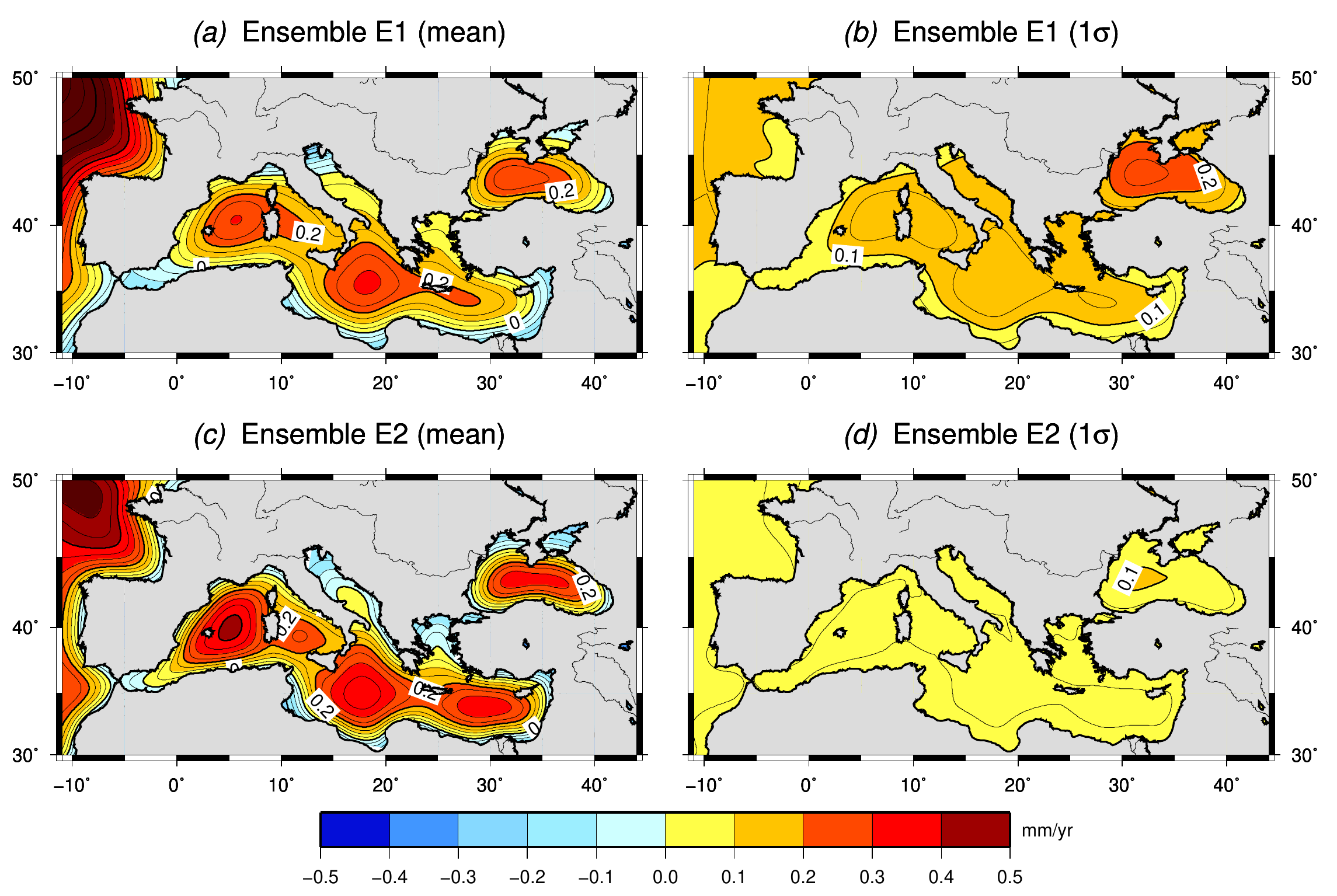}
\caption{Averages (left) and {standard deviations} (right) obtained from the GIA ensembles E1 (based upon ANU\_S4(E1)) and E2 (based upon ICE-6G$\_$C~(VM5a)) for the present-day rate of \sealevel change
across the Mediterranean region. Standard deviations {are relative to the average models} shown in the left frames.}
\label{figure:ensembles}
\end{figure}

In Figure~\ref{figure:ensembles} 
the average $\dot S_{avg}$ and the {standard deviation} $\sigma_{\dot S}$ of E1 and E2 are shown, for the current rate of relative \sealevel change. 
At a given location of coordinates $(\theta,\lambda)$, the standard deviation is $\sigma_{\dot S} = \sqrt{\sum_{i=1}^n (\dot S_i - \dot S_{avg} )^2/(n-1) }$, where $n$
is the number of samples in the ensemble.  
The general patterns are broadly comparable, and both maps still confirm the existence of late-Holocene high-stands and 
substantially agree on their former position. However, some significant qualitative differences can be evidenced: 
\emph{i)} sub-basin peak values for E2 generally exceed those based upon E1, \emph{ii)} the E2 
map is characterized by steeper gradients compared to E1, especially near the coastlines, \emph{iii)} the 
uncertainty (1-$\sigma$) associated to E1 exceeds that of E2. 

\begin{figure}
\centering
\includegraphics[width=0.7\textwidth,angle=0]{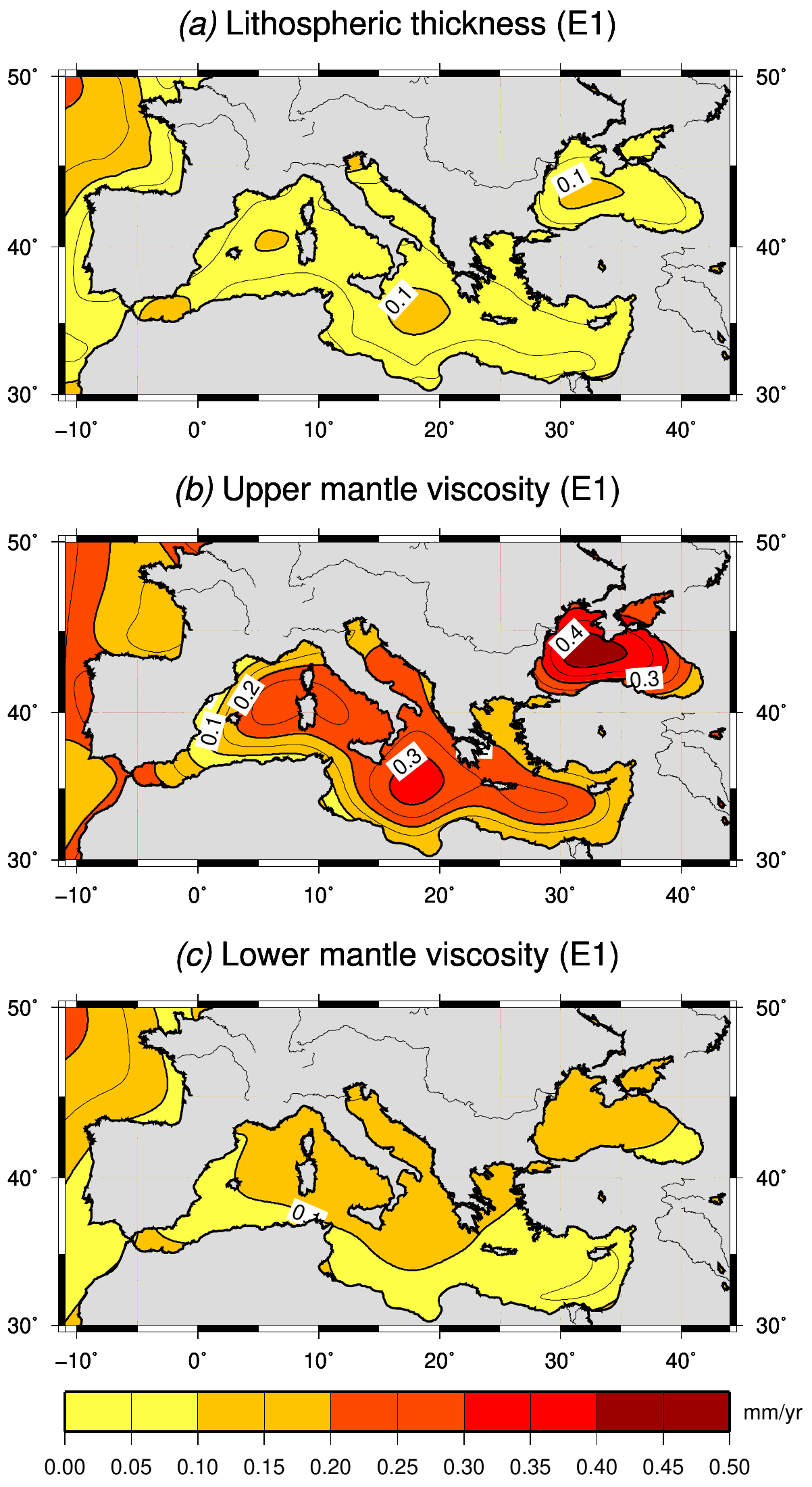}
\caption{{Standard deviation} of $\dot S$, evaluated varying individual rheological parameters, for the ensemble E1, based upon model ANU\_S4(E1).
Standard deviations are relative to the nominal model {ANU\_S4(E1)}.}
\label{fig:anu-ensemble}
\end{figure}

\begin{figure}
\centering
\includegraphics[width=1\textwidth,angle=0]{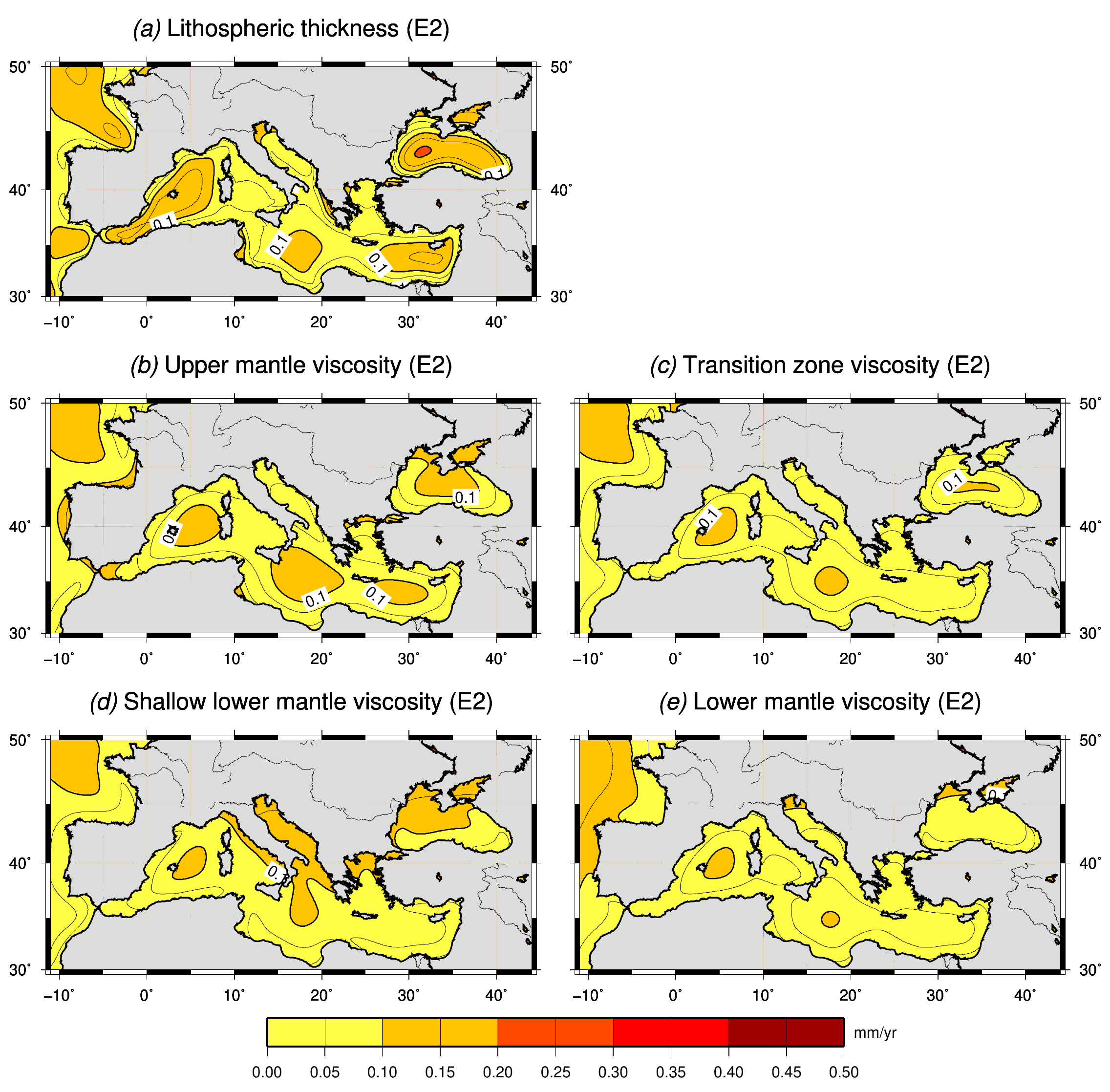}
\caption{{Standard deviation} of $\dot S$, evaluated varying individual rheological parameters, for the ensemble E2, constructed 
using model ICE-6G$\_$C~(VM5a). Standard deviations are relative to the nominal model ICE-6G$\_$C~(VM5a). 
~{}}
\label{fig:i6g-ensemble}
\end{figure}

To gain a better insight into the different patterns obtained in Figure~\ref{figure:ensembles}, 
in Figures~\ref{fig:anu-ensemble} and~\ref{fig:i6g-ensemble} we 
consider in detail the sensitivity of $\dot S$  to individual 
rheological parameters. More specifically, each of the panels show the {standard deviations} 
of the $\dot{S}$ map obtained by varying each parameter
within the range listed in Table~\ref{tab:ens}. The {standard deviations} are relative 
to the prediction ${\dot S}_{nom}$ obtained for the nominal model, which is 
defined by the values given in parentheses
in Table~\ref{tab:ens}. At a given position, the {standard deviations} have been computed by 
$\sigma_{\dot S} = \sqrt{\sum_{i=1}^n ( {\dot S}_i - {\dot S}_{nom} )^2/n}$,
where $n$ is the number of models in the sub-ensemble corresponding to variations of the chosen parameter. 
For the ensemble E1, sensitivity to the lithospheric thickness 
(Figure~\ref{fig:anu-ensemble}a) reaches a peak level of 
$0.1$ \rate at the center of the Ionian,
Balearic and Black seas, while it is generally minimum near the
shorelines. This pattern, which is correlated to the regional imprint of
$\dot{S}$ (see Figure~\ref{figure:peltier+lambeck}), hints to the role of lithospheric
flexure due to the meltwater load. Sensitivity to the upper mantle
viscosity (Figure~\ref{fig:anu-ensemble}{b}) is found to be considerably enhanced, with peak standard deviations 
in the range between $0.2$ and $0.4$ \rate in the central regions of the
sub-basins. As for the lithospheric thickness, the lowest sensitivity is
generally found across the coastlines, with the notable exception of the
Adriatic and Black seas. The sensitivity to the lower mantle viscosity (Figure~\ref{fig:anu-ensemble}c)
shows a completely different regional imprint, with standard deviations
increasing from the North African coast towards the northern margin
of the basin. {This pattern suggests an effect of the viscosity at depth
on the shape of the fore-bulge of the northern Europe ice sheets, as
pointed out by~\citet{royandpeltier2015} for the North American ice complex}.

Figure~\ref{fig:i6g-ensemble} shows the sensitivity to the rheology of present-day 
$\dot{S}$ predicted by the ICE-6G~(VM5a) GIA model, obtained through the
analysis of the E2 ensemble. Here, in addition to the
thickness of the elastic lithosphere, the viscosities of each of the
four layers assumed by the VM5a model have been independently explored.
For each parameter, the  pattern of the standard deviations appears correlated to the
spatial variability of $\dot{S}$ (see Figure~\ref{figure:peltier+lambeck}), with peak values
found in the central areas of the sub-basins. The sensitivity to
lithospheric thickness (Figure~\ref{fig:i6g-ensemble}a) turns out to be higher than in
ANU$\_$S4, with standard deviations of $0.2$ \rate found in the Balearic,
Levantine and Black seas. Conversely, the sensitivity to the upper
mantle viscosity structure (b and c) is much smaller than in
ANU$\_$S4, with standard deviations values not exceeding the $0.1$ \rate level both for
the upper mantle and the transition zone viscosities. A similar pattern
is found for the sensitivity to the lower mantle viscosity (d),
while for the viscosity of the shallow part of the lower mantle 
(e) the general imprint correlated with $\dot{S}$ is superimposed to a 
northward-increasing gradient, with peak values of $0.2$ \rate in the
northern Adriatic and across the {Crimean Peninsula}. 
{We note that, while the range of variability for the lithospheric thickness and lower mantle viscosity are considerably different between the two ensembles, all other viscosities are varied within comparable ranges of $0.7$--$0.8$ log Pa$\cdot$s.}
The  reduced sensitivity 
to the upper mantle viscosity profile shown by ICE-6G
may be related to the smaller viscosity contrasts assumed by VM5a with respect to ANU$\_$S4; 
{on the other hand, the pattern of the standard deviations
for the shallow lower mantle viscosity hints, as for ANU$\_$S4, to the
imprint of the lateral fore-bulge of the {northern European} ice complexes}.

\section{Discussion}\label{sec:discussion}

Previous work has clearly indicated that a correct interpretation of a number of geodetic signals, 
either regional or global, requires the proper evaluation of uncertainties associated with GIA modelling 
\citep{king2010improved,sasgen2012antarctic,caron2018gia,melini2019someremarks,li2020uncertainties}. 
For the first time, in this work we have explored in detail the features of various GIA signals across the Mediterranean
Sea, creating ensembles based upon up-to-date models recently proposed in the literature. Our study 
has been essentially motivated by the considerable amount of data indicating the variability of  
sea level in the past~\citep{sivan2001holocene,antonioli2009holocene,lambeck-purcell-2005,vacchi2016multiproxy} 
and the by efforts made to constrain the current deformations and \sealevel variability by high-quality geodetic
observations \citep{fenoglio2002long,serpelloni2013vertical,bonaduce2016sea}. Furthermore, the quality of existing data 
calls for a thorough evaluation of GIA modelling, improving upon previous works in which now outdated 
GIA models have been used or some physical ingredients of the SLE have been {not taken into account} 
\citep[see \emph{e.g.,}][]{stocchi2009influence}. Assuming a spherically symmetric Earth structure with a 
linear viscoelastic rheology, two ensembles of 
GIA models have been proposed, starting from structurally different nominal models developed by the two independent 
schools led by Kurt Lambeck (Australian National University) and {WR Peltier} (Univ. of Toronto), respectively. 
Overall, the two ensembles encompass $\sim 120$ GIA models, a small number compared to the GIA ensemble
of \citet{caron2018gia} but largely improving the ``mini-ensemble'' approach of \citet{melini2019someremarks}. 
The range of viscosity profiles that we have adopted has been
suggested by previous works by the two groups (see \citeauthor{lambeck-purcell-2005}~\citeyear{lambeck-purcell-2005} 
and \citeauthor{royandpeltier2015}~\citeyear{royandpeltier2015}, respectively).  Although limited to $1$-D Earth's structures, 
our results clearly {suggest} that a high-resolution approach is necessary in order to capture the 
small-scale details of the ongoing GIA contributions across the study region, which are characterized by a striking 
regional variability despite the relatively narrow region. This variability is particularly enhanced 
for the fields $\dot S$ and $\dot U$, while $\dot N$, $\dot U_n$ and $\dot U_e$ are characterized by a comparatively 
smoother pattern although some of them have a significant amplitude. 

\begin{figure}
\centering
\includegraphics[width=0.8\textwidth,angle=0]{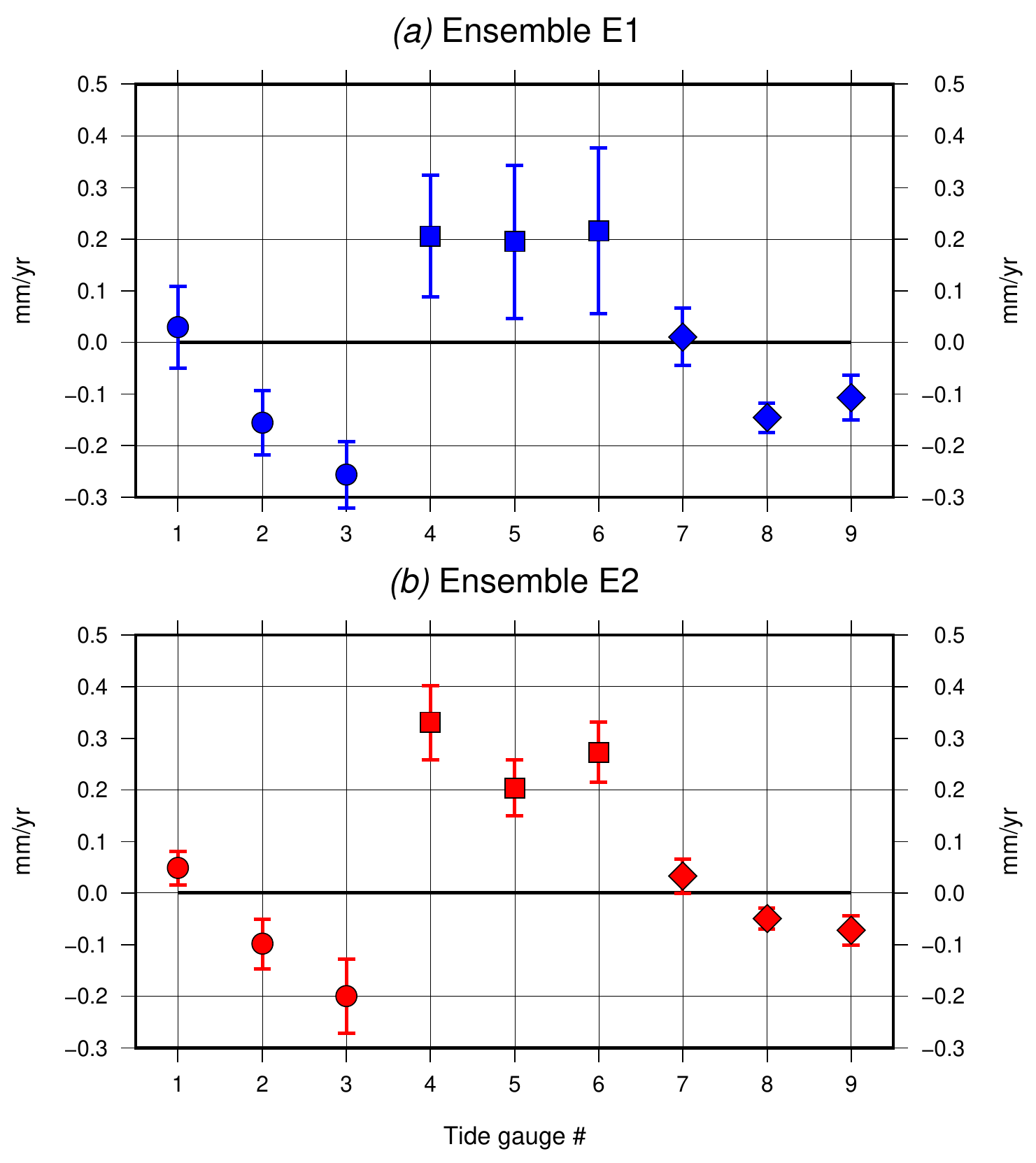}
\caption{Ensemble average predictions of $\dot S$ at Mediterranean tide gauges and their $1$-$\sigma$ uncertainties, 
according to 
ensemble E1 (top) and E2 (bottom). The locations of the tide gauges {are} {} shown in Figure~\ref{figure:peltier+lambeck}. Standard deviations 
are relative to the ensemble average.
{Numerical values are listed in Table \ref{tab:rates}}.}
\label{fig:tg-ensemble}
\end{figure}

In previous investigations aimed at studying the \sealevel variability in the Mediterranean Sea 
\citep[see \emph{e.g.}][]{bonaduce2016sea} the effects of GIA have been estimated using a single model and 
focussing on the tide gauge locations, without considering the pattern of the GIA imprint across the 
whole region and neglecting the possible uncertainties involved in modelling. 
In other studies \citep[see \emph{e.g.}][]{santamaria2017uncertainty} the uncertainty has been roughly estimated
by considering the predictions by two independently developed GIA models, basically adopting
the ``mini-ensemble'' approach followed by \citet{melini2019someremarks}. 
In Figure~\ref{fig:tg-ensemble} we consider the same nine tide gauges whose locations are marked by 
green symbols in Figure~\ref{figure:peltier+lambeck}. For each of them we show 
the average GIA contribution to the rate of relative \sealevel change and its $1$-$\sigma$ uncertainty estimated by the two ensembles 
E1 (top) and E2 (bottom) introduced in previous section. 
For ensemble E1, it clearly appears that tide gauge sites 
located in islands in the bulk of the Mediterranean basin 
(sites $4$, $5$ and $6$) are characterized by the larger GIA rates, but also by the largest uncertainties. Conversely, 
in places located along the continental coastlines the GIA rates are comparatively modest, and generally less 
uncertain. This trade-off is less evident when we consider the ensemble E2, which is generally characterized by 
smaller uncertainties compared with E1. The results of Figure~\ref{fig:tg-ensemble}, along with the 
patterns shown in Figure~\ref{figure:ensembles}, clearly indicate that the $\dot S$ values expected at tide gauges 
fall on the range between $-0.3$ and $0.5$ \rates, values that are smaller than those predicted in formerly
de-glaciated areas by one order of magnitude \citep[see \emph{e.g.,}][]{melini2019someremarks}, regardless of the ensemble considered.  
Although the $1$-$\sigma$ uncertainties show a significant regional variability across the Mediterranean basin, our results suggest that 
they may attain a maximum value of $0.2$ \rates. 

\begin{figure}
\centering
\includegraphics[width=0.8\textwidth,angle=0]{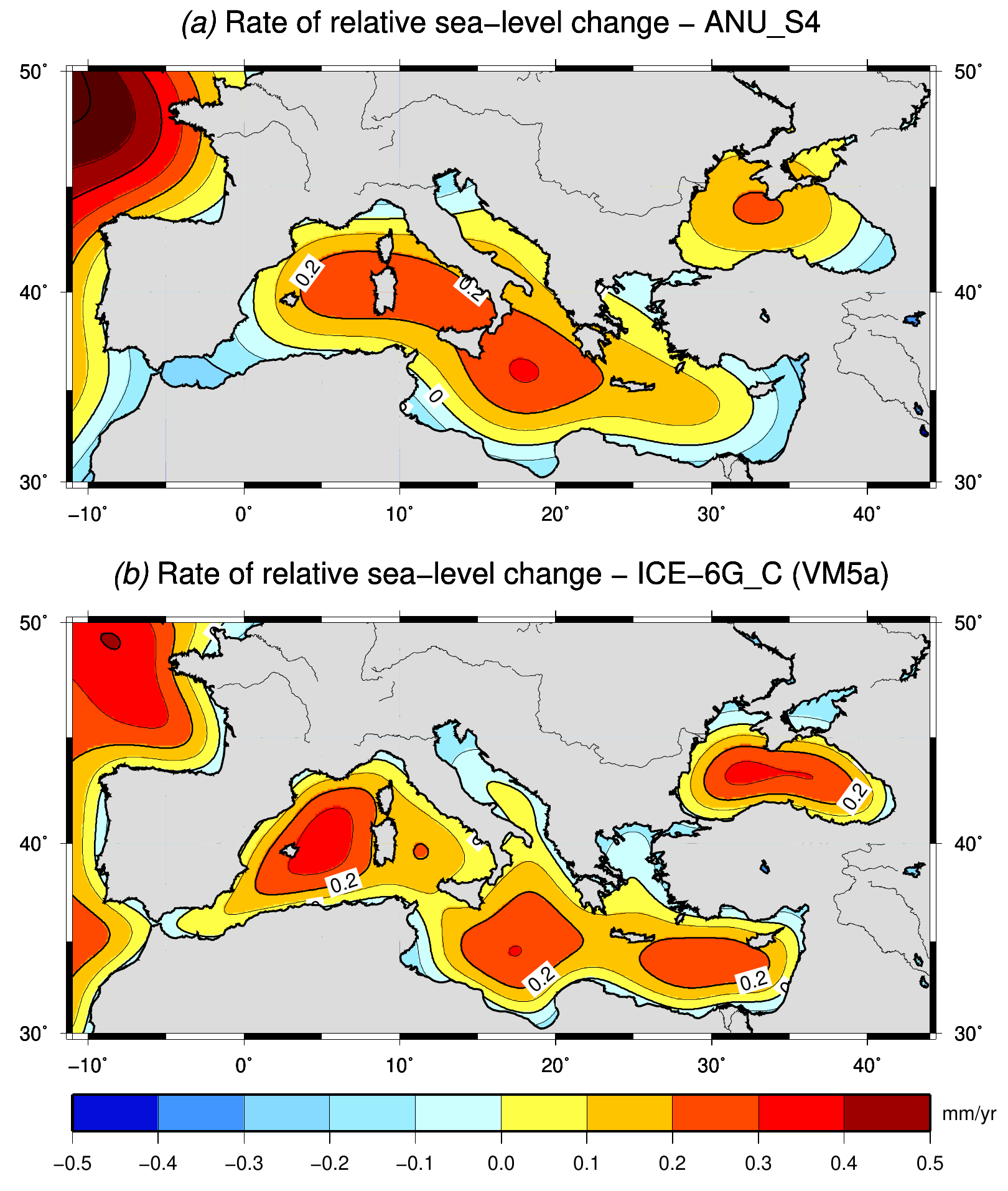}
\caption{GIA rates of relative \sealevel change ($\dot S$) expected in two thousands years across the Mediterranean Sea, according to the 
two models ANU$\_$S4 (a) and ICE-6G$\_$C~(VM5a) (b).}
\label{fig:future}
\end{figure}

Up to now, our attention has been limited to the ongoing geodetic variations due to GIA. In GIA modelling, these variations
are generally assumed to be constant on a century time scale, since it is expected that the relatively high average mantle 
viscosity would prevent a significant decay \citep[see \emph{e.g.,}][]{galassi2014sea,spada2016spectral}. To test this hypothesis, in  Figure~\ref{fig:future} we have projected the 
$\dot S$ GIA imprints over the next two millennia, using 
{the two models}
ANU$\_$S4 and 
ICE-6G$\_$C~(VM5a)  
whose contemporary imprints have been considered already in Figure~\ref{figure:peltier+lambeck}. It is useful to remark that
according to these two models (see Table~\ref{tab:ens}), the bulk viscosity of the mantle is not too dissimilar from the ``Haskell value'' 
of $10^{21}$ Pa $\cdot$ s \citep[see][and references therein]{mitrovica1996haskell}, so that a relaxation time of a few millennia would be 
expected~\citep[see \emph{e.g.}][]{turcotte2014geodynamics}. Indeed, the two figures show 
that the GIA rates are expected to decrease considerably (by $\sim 50\%$) over the next two millennia, 
leaving the general pattern of $\dot S$ basically unmodified with respect to the current imprint. At the same time,
we have verified that the rates would be effectively unchanged over one century across the whole Mediterranean
Basin. We note, however, {that} the nearshore mitigating effect of GIA ($\dot S<0$) will yet persist at some specific locations, {although} it is not expected 
that these relatively small rates (a few fractions of millimeters per year) would significantly counteract the general rising trend of \sealevel due 
to climate change over next millennia~\citep[see \emph{e.g.,}][]{slangen2012towards}. 

\section{Conclusion}\label{sec:conclusions} 

In this work, we have obtained a set of high-resolution numerical
solutions for the SLE at the Mediterranean scale, based on two
up-to-date GIA models independently developed by the research groups led
by Kurt Lambeck and WR Peltier. We have shown that some spatial features
of the $\dot{S}$ maps are remarkably common between the two GIA models:
\textit{i)}: peak values of $\dot{S}$, of about $0.4$ \rates, are attained
in the central sub-basins, hinting to the effect of lithospheric flexure
in reponse to hydrostatic load; \textit{ii)} swaths of \sealevel
\textit{fall} ($\dot{S}<0$) are present in narrow coastal inlets,
implying the possible existence of Holocene high-stands, suggested by some works in the 
literature.  
Rates of horizontal displacements due
to ongoing isostatic readjustment point to a nearly uniform northward drift across
the Mediterranean, associated with the collapse of the {northern European} ice
sheet fore-bulges. 
{With an amplitude of $\sim 0.8$ \rates, this northward drift notably exceeds the vertical rates.} 

Through an ensemble approach, we assessed the model uncertainties
associated to present-day GIA across the Mediterranean basin.
Uncertainties on $\dot{S}$ are generally correlated with the amplitude
of the field and reach the $0.1$ \rate level in the central sub-basins.
A remarkable exception is seen for 
{the ensemble based on ANU$\_$S4},
for which a
sensitivity to the viscosity structure in the upper mantle reaches the
$0.4$ \rate level, 
possibly associated to the enhanced viscosity contrasts assumed by ANU$\_$S4 {and to the correspondingly weaker upper mantle viscosities explored in ensemble E1}.
{In both ensembles, the spatial pattern of sensitivity
to the lower mantle viscosity presents a north-south trending gradient, hinting
to a long-wavelength effect associated with the isostatic response to
the melting of Fennoscandian ice complexes.}
{For a more comprehensive assessment of the GIA modeling uncertainties in the Mediterranean Sea, the 
regional-scale geodynamical setting would need to be taken into consideration. However, 
the structural complexity of region can only be accounted for through 
a 3-D numerical approach, which is far beyond the reach of the class of
GIA models considered in this study.}

\section*{Acknowledgments}
{We thank Ga{\"e}l Choblet and an anonymous reviewer for very constructive comments, and Matteo Vacchi for discussion and encouragement}. GS is funded by a FFABR (Finanziamento delle 
Attivit\`a Base di Ricerca) grant of MIUR (Ministero
dell’ Istruzione, dell’Università e della Ricerca) and by a research
grant of DIFA (Dipartimento di Fisica e Astronomia ``Augusto Righi'') {of the Alma Mater Studiorum Università di Bologna}. DM is 
funded by a INGV (Istituto Nazionale di Geofisica e Vulcanologia) ``Ricerca libera 2019'' research grant and by the INGV 
MACMAP departmental project.

\section*{Data availability}
{The data underlying this article will be shared on reasonable request to the corresponding author.}

\clearpage

\clearpage

\clearpage

\clearpage

\clearpage

\clearpage
\begin{table}
\caption{{Range of variability for rheological parameters in ensembles E1 and E2, based on nominal models ANU\_S4(E1) and
ICE-6G~(VM5a), respectively. Note that models in ensemble E1 do not include transition zone and shallow lower mantle layers.
All models in ensemble E2 assume
a $40$ km-thick
layer with viscosity $10^{22}\,\mathrm{Pa}\cdot\mathrm{s}$ at the base of the elastic lithosphere.
For each 
ensemble,
nominal values of parameters are shown in parentheses.}} 
\begin{center}
\begin{tabular}{lcccc}
\hline 
Ensemble  & \multicolumn{2}{c}{E1} & \multicolumn{2}{c}{E2} \\
Nominal model  & \multicolumn{2}{c}{ANU\_S4(E1)} & \multicolumn{2}{c}{ICE-6G~(VM5a)} \\
\hline 
Lithosphere thickness (km) &                    $45\div 85$     & (65) &
$60\div 200$ & (100)\\
\hline
Upper mantle viscosity (log, Pa$\cdot$s)    &  $19.9\div 20.7$ & (20.5)&
$20.2 \div 20.9$ & (20.7)\\
Transition zone viscosity (log, Pa$\cdot$s) &  --- &           &
$20.5\div21.2$ & (20.7)\\
\hline
Shallow lower mantle viscosity (log, Pa$\cdot$s) & --- &       &
$20.7\div21.5$ & (21.2)\\
Lower mantle viscosity (log, Pa$\cdot$s) & $21.4\div 22.8$ & (22.0)  &
  $21.3\div22.0$ & (21.5)
\\
\hline
\end{tabular}
\end{center}
\label{tab:ens}
\end{table}

\clearpage

\clearpage

\clearpage

\clearpage

\clearpage
\begin{table}
\begin{center}
\caption{
{Rates of relative sea-level change expected at tide gauges according to models ANU\_S4 and ICE-6G\_C(VM5a), and corresponding averages and 1-$\sigma$ uncertainties according to ensembles E1 and E2. ANU\_S4(E1) corresponds to the nominal model of \citet{lambeck-purcell-2005}, upon which Ensemble E1 is based, and it differs from ANU\_S4 for the upper mantle viscosity and lithospheric thickness. All rates are in units of \rates.}
}\label{tab:rates}
\begin{tabular}{lccccc}
\hline
Tide gauge  & \multicolumn{1}{c}{ANU\_S4} & \multicolumn{1}{c}{ANU\_S4(E1)} & \multicolumn{1}{c}{E1 Ensemble} & \multicolumn{1}{c}{ICE-6G} & \multicolumn{1}{c}{E2 Ensemble}\\
\hline
1. Marseilles &  $+0.08$ &  $+0.03$ &  $+0.03 \pm 0.08$     &  $+0.05$  & $+0.05 \pm 0.03$\\
2. Genova     & $-0.15$ & $-0.17$ &  $-0.16 \pm 0.06$     & $-0.10$  & $-0.10 \pm 0.05$\\
3. Trieste    & $-0.20$ & $-0.27$ &  $-0.26 \pm 0.06$     & $-0.21$  & $-0.20 \pm 0.07$\\
\hline
4. Palma de Mallorca & $+0.28$ & $+0.22$  & $+0.21 \pm 0.12$  &   $+0.33$  &  $+0.33 \pm 0.07$ \\
5. Cagliari          & $+0.38$ & $+0.23$  & $+0.20 \pm 0.15$  &   $+0.20$  &  $+0.20 \pm 0.05$ \\
6. Valletta          & $+0.33$ & $+0.26$  & $+0.22 \pm 0.16$  &   $+0.27$  &  $+0.27 \pm 0.06$ \\
\hline
7. Alicante I       & $-0.03$ & $-0.01$  & $+0.01 \pm 0.06$ &  $+0.03$  &  $+0.03 \pm 0.03$ \\
8. Hadera           & $-0.19$ & $-0.15$  & $-0.15 \pm 0.03$ & $-0.05$  & $-0.05 \pm 0.02$ \\
9. Alexandria       & $-0.08$ & $-0.10$  & $-0.11 \pm 0.04$ & $-0.07$  & $-0.07 \pm 0.03$ \\
\hline
\end{tabular}
\end{center}
\end{table}

\clearpage

\clearpage
\bibliographystyle{gji}
\bibliography{scibib-GIORGIO.bib}

\end{document}